# ACCURACY OF MRI CLASSIFICATION ALGORITHMS IN A TERTIARY MEMORY CENTER CLINICAL ROUTINE COHORT

Alexandre Morin, MD [1, 2, 3,*], Jorge Samper-Gonzalez [2, 3], Anne Bertrand, MD, PhD [2, 3, 4, §], Sébastian Ströer, MD [2, 4], Didier Dormont, MD, PhD [2, 3, 4], Aline Mendes, MD [5], Pierrick Coupé, PhD [6], Jamila Ahdidan, PhD [7], Marcel Lévy, MD [5], Dalila Samri [5], Harald Hampel, MD, PhD [5, 8, 9], Bruno Dubois, MD [5, 9], Marc Teichmann, MD, PhD [5, 9], Stéphane Epelbaum, MD, PhD [2, 3, 5] and Olivier Colliot, PhD [2, 3, 4, 5 *]

[1] AP-HP, Hôpital de la Pitié-Salpêtrière, Department of Neurology, Unité de Neuro-Psychiatrie Comportementale (UNPC), F-75013, Paris, France.
[2] Sorbonne Universités, UPMC Univ Paris 06, Inserm, CNRS, ICM, F-75013 Paris, France
[3] Inria, Aramis-project team, Paris, France
[4] AP-HP, Hôpital de la Pitié-Salpêtrière, Department of Neuroradiology, F-75013, Paris, France
[5] AP-HP, Hôpital de la Pitié-Salpêtrière, Department of Neurology, Institut de la Mémoire et de la Maladie d'Alzheimer (IM2A), F-75013, Paris, France
[6] Laboratoire Bordelais de Recherche en Informatique, Unit Mixte de Recherche CNRS (UMR 5800), PICTURA Research Group, Bordeaux, France
[7] Brainreader, Horsens, Denmark
[8] AXA Research Fund & UPMC Chair, Paris, France; Sorbonne Universities, Pierre et Marie Curie University, Paris 06,
[9] ICM, ICM-INSERM 1127, FrontLab.

§ Deceased, March 2nd, 2018

*Correspondence to:
Olivier Colliot - olivier.colliot@upmc.fr
ICM – Brain and Spinal Cord Institute
ARAMIS team
Pitié-Salpêtrière Hospital
47-83, boulevard de l'Hôpital, 75651 Paris Cedex 13, France
Phone : 01 57 27 43 65

# Abstract

**Background:** Automated volumetry software (AVS) has recently become widely available to neuroradiologists. MRI volumetry with AVS may support the diagnosis of dementias by identifying regional atrophy. Moreover, automatic classifiers using machine learning techniques have recently emerged as promising approaches to assist diagnosis. However, the performance of both AVS and automatic classifiers has been evaluated mostly in the artificial setting of research datasets.

**Objective:** Our aim was to evaluate the performance of 2 AVS and an automatic classifier in the clinical routine condition of a memory clinic.

**Methods:** We studied 239 patients with cognitive troubles from a single memory center cohort. Using clinical routine T1-weighted MRI, we evaluated the classification performance of: i) univariate volumetry using two AVS (volBrain and Neuroreader™); ii) Support Vector Machine (SVM) automatic classifier, using either the AVS volumes (SVM-AVS), or whole gray matter (SVM-WGM); iii) reading by two neuroradiologists. The performance measure was the balanced diagnostic accuracy. The reference standard was consensus diagnosis by three neurologists using clinical, biological (cerebrospinal fluid) and imaging data and following international criteria.

**Results:** Univariate AVS volumetry provided only moderate accuracies (46% to 71% with hippocampal volume). The accuracy improved when using SVM-AVS classifier (52% to 85%), becoming close to that of SVM-WGM (52 to 90%). Visual classification by neuroradiologists ranged between SVM-AVS and SVM-WGM.

**Conclusion:** In the routine practice of a memory clinic, the use of volumetric measures provided by AVS yields only moderate accuracy. Automatic classifiers can improve accuracy and could be a useful tool to assist diagnosis.

**Acknowledgements**

O.C. is supported by a "Contrat d'Interface Local" from Assistance Publique-Hôpitaux de Paris (AP-HP). HH is supported by the AXA Research Fund, the Fondation Université Pierre et Marie Curie and the Fondation pour la Recherche sur Alzheimer, Paris, France.

**Source of funding**

The research leading to these results has received funding from the French government under management of Agence Nationale de la Recherche as part of the "Investissements d'avenir" program, reference ANR-19-P3IA-0001 (PRAIRIE 3IA Institute), reference ANR-10-IAIHU-06 (Agence Nationale de la Recherche-10-IA Institut Hospitalo-Universitaire-6), and reference ANR-11-IDEX-004 (Agence Nationale de la Recherche-11- Initiative d'Excellence-004, project LearnPETMR number SU-16-R-EMR-16), from the European Union H2020 program (project EuroPOND, grant number 666992), from the joint NSF/NIH/ANR program "Collaborative Research in Computational Neuroscience" (project HIPLAY7, grant number ANR-16-NEUC-0001-01), from Agence Nationale de la Recherche (project PREVDEMALS, grant number ANR-14-CE15-0016-07), from the ICM Big Brain Theory Program (project DYNAMO), from the Abeona Foundation (project Brain@Scale) and from the "Contrat d'Interface Local" program (to Dr Colliot) from Assistance Publique-Hôpitaux de Paris (AP-HP).

**Role of the funding source**

The sponsors of the study had no role in study design, data analysis or interpretation, writing or decision to submit the report for publication.

# Introduction

**Background:**

The diagnostic criteria of Alzheimer's disease (AD) and other dementias have evolved in the past decades from a clinical descriptive perspective to biomarker-supported definitions, mainly due to innovation in brain imaging, and biological fluid markers [1]. Among neuroimaging biomarkers, MRI is the less invasive, most widely available, cost-effective, is systematically recommended in dementia and can provide supportive criteria for many neurodegenerative conditions [2–4]. MRI can identify areas of atrophy that can suggest particular types of dementia, such as atrophy of the medial temporal structures in late-onset AD [5,6] or anterior atrophy in frontotemporal dementia [7]. Assessment of regional atrophy using MRI in dementia has been extensively studied using visual, semi-quantitative ratings [5–7], manual volumetry, and more recently Automated Volumetry Software (AVS)[8–11].

AVS such as Neuroreader™ [10], and volBrain [12] provide volumetric measures of anatomical structures. Unlike subjective visual analysis of atrophy, AVS provide objective, quantitative measurement of various regions of interest (ROI) volumes. These tools, which are progressively being implemented in clinical MRI software have only been evaluated in research settings [10,11,13,14]. Besides, due to their univariate nature, they cannot detect complex multivariate combinations of regional atrophies, essential to discriminate between different dementias.

Automatic classifiers, based on machine learning techniques, are able to automatically learn complex multivariate discriminative patterns without priors on specific anatomical structures. Automatic classifiers have also mainly been evaluated in research settings, with standardized MRI acquisition and focusing on a single type of dementia (most often Alzheimer's disease) and age-matched healthy controls [15–19].

**Objective:** In this study, we evaluated the diagnostic classification performance of AVS volumetry (volBrain and Neuroreader™), automatic classifiers (based on whole gray matter or on AVS volumes), in a clinical routine cohort of patients presenting with various neurodegenerative dementia disorders, depression or subjective cognitive decline.

# Material and Methods

## Participants

All subjects were recruited retrospectively in a tertiary academic expert memory center (Institute for Memory and Alzheimer's disease – Department of Neurology, Pitié-Salpêtrière University Hospital) from the ClinAD cohort [20]. The ClinAD cohort consists of 992 consecutive patients who consulted from 2005 to 2014 for cognitive impairment and who underwent lumbar puncture. Data collection was planned before the index test and reference standard were performed. All patients had neurological, biological and neuropsychological evaluations. Cerebrospinal fluid (CSF) $A\beta_{1-42}$, tau and phosphorylated tau was available for all participants. All clinical and biological data were generated during a routine clinical workup and were retrospectively extracted for the purpose of this study. Therefore, according to French legislation, explicit consent was waived. However, regulations concerning electronic filing were followed, and patients and their relatives were informed that anonymised data might be used in research investigations.

For each patient, the diagnosis was assessed by a group of 3 neurologists based on clinical, biological and imaging data, following international consensus criteria for AD (IWG-2) [21], fronto-temporal dementia (FTD) [2], primary progressive aphasia (PPA) of the logopenic (lv-PPA), semantic (SD) or non-fluent/agrammatic (nf-PPA) [22] variant, cortico-basal syndrome (CBD) [4], progressive supranuclear palsy (PSP) [23], posterior cortical atrophy (PCA) [24], Lewy body dementia (LBD) [25], and depression [26]. This consensus diagnosis formed the reference standard. The classifier and volumetry (index tests) results were not available to assessors of the reference standard. As clinical presentations and atrophy patterns depend mostly on the age of onset of AD [27], the AD group was separated into Early-onset AD (EOAD) and Late-Onset-AD (LOAD), with age of onset respectively before and after 65 years. In addition, 342 out of 992 patients were excluded because they presented with mixed pathology, vascular disease (Fazekas score > 2 or significant stroke) or unclear diagnosis. From the 650 patients of the ClinAD cohort, 380 patients were excluded because the MRI was performed outside our center and was not available for our study, resulting in 270 patients. We added 12 subjective cognitive decline (SCD) patients, defined as patients with cognitive complaint but with normal neuropsychological examination.

Among the 282 patients, 7 were excluded due to poor image quality or failure of image processing pipelines. Specifically, 6 had a very low MRI quality on visual analysis (missing slices or strong motion artifacts) and the image processing pipelines failed in one participant. The quality of the remaining MRI data was variable, reflecting the reality of clinical routine, but proved sufficient for reliable image processing. The quality of image segmentation results was visually assessed. Moreover, we excluded diagnostic groups with less than 15 patients (nf-PPA, PSP, PCA) as automatic classifiers cannot be trained robustly on very small groups of subjects. As a result, the analyses were performed on 239 patients belonging to the following eight diagnostic groups: cortico-basal syndrome, early-onset AD, late-onset AD, fronto-temporal dementia of the behavioral type, Lewy body dementia, logopenic variant of primary progressive aphasia, semantic variant of primary progressive aphasia, and depression. The flow chart is described on Supplementary table 1. In this cohort, the only group without degenerative condition was that of patients with depression. We aim to compare the results obtained for depression to that obtained for subjective cognitive decline (SCD). To that purpose, we added 12 patients with SCD, defined as patients with cognitive complaint but with normal neuropsychological examination. For this group, classifiers were trained using the depression group and applied to the SCD group, because the training of the classifier on 12 participants would not be robust enough.

Demographic data are summarized in Table 1. Difference between groups on demographic and clinical data was evaluated with ANOVA for continuous data and $X^2$ test for binary data using XLStat Software (Addinsoft, www.xlstat.com). As expected, since we separated the AD group in LOAD and EOAD, age at diagnosis was significantly different between groups (in ANOVA and Post-Hoc Test). The MMSE score was also different since the neurodegenerative conditions do not have the same cognitive profile. For example, language impairment in PPA usually leads to lower MMSE scores than frontal dysfunction in FTD. There was no difference between groups regarding gender and MRI magnetic field.

**MRI Acquisition**

All 239 patients had an available brain MRI performed in the Department of Neuroradiology at Pitié-Salpêtrière Hospital: 63 on a 3T MRI GE Sigma HD, 9 on a 1.5 T MRI GE Optima 450, 44 on a 1.5T MRI

GE Signa Excite and 123 on a 1T MRI Philips Panorama. All MRI included a 3D T1-weighted sequence with a spatial resolution ranging from 0.5x0.5x1.2mm3 to 1x1x1.2mm3. Since imaging was performed as part of clinical routine, MRI acquisition parameters were not homogenized. Sequence parameters are available in Supplementary Table 2. The 12 SMC patients had an MRI performed in our center: 8 on a 3T MRI GE Signa HD, 1 on a 1.5 T MRI GE Optima 450, and 3 on a 1T MRI Philips Panorama.

## Fully Automated Volumetry Software

The Neuroreader™ software (http://www.brainreader.net) is a commercial clinical brain image analysis tool [10]. The system provides the volumes of the following structures: intracranial cavity, tissue categories (WM, GM, and CSF), subcortical GM structures (putamen, caudate, pallidum, thalamus, hippocampus, amygdala and accumbens) and lobes (occipital, parietal, frontal and temporal). Processing times range from 3 to 7 minutes as a function of image size, irrespective of magnetic field strength.

The volBrain software (http://volBrain.upv.es) is an online freely-available academic brain image analysis tool [12]. The volBrain system takes around 15 minutes to perform the full analysis and provides the same volumes as Neuroreader™ except for the lobar volumes, only provided by Neuroreader™. However, the volBrain system provides hemisphere, brainstem and cerebellum segmentations which were not used in this study.

## Automatic classification using SVM

### Pre-Processing: extraction of Whole Grey Matter maps

All T1-weighted MRI images were segmented into Gray Matter (GM), White Matter (WM) and CSF tissues maps using the Statistical Parametric Mapping unified segmentation routine with the default parameters (SPM12, London, UK http://www.fil.ion.ucl.ac.uk/spm/) [28]. A population template was calculated from GM and WM tissue maps using the DARTEL diffeomorphic registration algorithm with the default parameters [29]. The obtained transformations and a spatial normalization were applied to the GM tissue maps. All maps were modulated to ensure that the overall tissue amount remains constant and normalized to

MNI space. 12mm smoothing was applied as the classification performed better with this parameter than with none or less smoothed images.

**SVM classification**

Whole Gray Matter (WGM) maps were then used as input of a high-dimensional classifier, based on a linear support vector machine (SVM) classifier. In brief, the linear SVM looks for a hyperplane which best separates two given groups of patients, in a very high dimensional space composed of all voxel values. In such approach, the machine learning algorithm automatically learns the spatial pattern (set of voxels and their weights) allowing to discriminate between diagnostic group. Importantly, the classifier does not use prior information such as anatomical boundaries between structures or that a specific anatomical structure (e.g. hippocampus) would be affected in a given condition. Please refer to Cuingnet et al. [15] for more details.

SVM classification was performed for each possible pair of diagnostic groups (e.g. EOAD vs. FTD, LOAD vs. FTD…). The performance measure was the balanced diagnostic accuracy defined as: (sensitivity – specificity)/2. Unlike standard accuracy, balanced accuracy allows to objectively compare the performance of different classification tasks even in the presence of unbalanced groups [15].

In order to compute unbiased estimates of classification performances, we used a 10-fold cross validation, meaning that each 10% of the set is used for testing and the other 90% for training, changing the groups in each out of the ten trials. This ensures that the patient that is currently being classified has not been used to train the classifier, a problem known as "double-dipping". Finally, the SVM classifier has one hyper-parameter to optimize. The optimization was done using a grid-search. Again, in order to have a fully unbiased evaluation, the hyper-parameter tuning was done using a second, nested, 10-fold cross-validation procedure.

Finally, in order to have a fair comparison between WGM maps and AVS volumes, we also performed SVM classification using volumes of each AVS as input, all regional volumes (for a given AVS) being simultaneously used in a multivariate manner.

## Radiological classification

Two neuroradiologists (AB, with 8 years of experience, and SS, with 4 years of experience), specialized in the evaluation of dementia, performed a visual classification of three diagnosis pairs on the same dataset: FTD vs EOAD, depression vs LOAD and LBD vs LOAD. We chose FTD vs EOAD and depression vs LOAD for their relevance in clinical practice. We chose LBD vs LOAD because the SVM classifier yielded only moderate accuracies, and because the diagnosis of LBD based on MRI is difficult. The neuroradiologists were blind to all patient data except MRI.

# Results

## Automated Volumetry Software: volBrain and Neuroreader™

We performed a univariate classification based on each AVS volume separately. Volumes were normalized to the measured Total Intracranial volume (mTIV) (using the formula: Volume/mTIV), as discrimination was slightly better than with absolute values. VolBrain and Neuroreader™ performed similarly on univariate classification with balanced accuracy rates ranging from 46% to 71% based on hippocampal volumes. We show various volumes obtained in Neuroreader™ in Supplementary Table 3. We show results of classification based on hippocampal volume computed with Neuroreader™ in Table 2. In Supplementary Table 4 to 9, we provide classification balanced accuracy based on volumes of other anatomical structures, known to be of particular interest in various neurodegenerative conditions.

## Automatic SVM classifier from Whole Gray Matter maps

Table 3 provides the results of automatic SVM classification from WGM segmentation maps. Balanced accuracies ranged from 52% (LBD vs LOAD) to 90% (EarlyAD vs SCD). We present in Supplementary Figure 1 two examples of weight maps, which are graphic representations of the most relevant voxels for classification.

## Automatic SVM classification from AVS volumes

To fully compare AVS with our SVM-WGM classification, we provide, in Supplementary Table 10, results of SVM classification from all volumes obtained with volBrain and Neuroreader™ in addition to SVM based on WGM. In general, results were slightly lower than with SVM classification from WGM. Overall, volBrain and Neuroreader™ performed similarly, even though one or the other tool achieved slightly higher performances in some specific cases.

## Radiological classification

Classification by experienced neuroradiologists resulted in the following balanced accuracies : 77% (neuroradiologist 1) and 72% (neuroradiologist 2) for LOAD vs depression, 72% and 75% for FTD vs EOAD, 57% and 63% for LBD vs LOAD (Table 4). Neuroradiological classification performed better than both SVM-AVS and univariate AVS except for LBD vs LOAD classification in which they performed equally. The performance of the SVM-WGM was in general comparable to that of neuroradiologists. However, it was superior to both radiologists for FTD vs EOAD classification.

# Discussion

In this study, we assessed the diagnostic performance of AVS and SVM classifiers for various neurodegenerative conditions. SVM classifier based on whole gray matter provided accurate diagnostic classification for the majority of diagnoses and was far more accurate than univariate classification based on regional volumes such as hippocampal volume obtained through AVS. The performance of the SVM classifier was similar or slightly higher to that of trained neuroradiologists on selected classification tasks.

The best accuracies were obtained with SVM classification from whole gray matter maps. Balanced accuracy was superior to 70% in 64% of the available combinations and superior to 80% in 25% of them. Two studies evaluated SVM classification between AD and FTD in a research setting [16,30]. In this setting, they obtained slightly higher diagnostic classification, with AD vs. FTD classification ranging from 84% to 90% (in our study: FTD vs. EOAD: 83% and FTD vs. LOAD: 73%). This slightly superior accuracy might be explained by the more controlled setting of research studies, in particular less heterogeneous MRI acquisitions, and by the fact that our patients were at a slightly less advanced disease stage. Moreover, in Klöppel et al. [30], the use of anatomopathology as the diagnosis criteria, might have provided more homogeneous groups of patients, helping to better distinguish different diagnoses. To the best of our knowledge, only one study has previously evaluated SVM classifiers in clinical routine with various types of dementia [31]. The accuracies that we report are consistent with those reported in Koikkalainen et al, [31] in which diagnostic accuracy for FTD vs. AD was 80% (in our study, FTD vs. LOAD: 73% and FTD vs. EOAD: 83%), for LBD vs. AD 68% (in our study, LBD vs. EOAD: 77% and LBD vs. LOAD: 52%) and for LBD vs. FTD 77.5 (in our study, LBD vs. FTD: 67%). In this previous study, as compared to ours, there wasn't any patient with PPA or CBD. Furthermore, contrarily to our study, diagnoses were not assessed with the latest diagnosis criteria, especially regarding Alzheimer's CSF biomarkers. Finally, this study did not compare the performance of SVM to that of AVS tools which are quickly becoming standard in radiological routine. Therefore, to the best of our knowledge, we present the first study of whole-brain classifiers on clinical routine data based on the latest diagnostic criteria, and with comparison to AVS tools, the current standard of quantitative clinical radiology.

When focusing on some particularly difficult clinical situations, automatic classification results are particularly promising. For instance, SVM classification distinguished depression, EOAD and FTD with an accuracy superior to 80%. In particular, SVM classification was more accurate than that of trained neuroradiologists for EOAD vs FTD. These situations often imply facing young patients, with an atypical symptomatic presentation. In these cases, there is often a dramatic impact on the professional and familial life. Finally, the diagnosis implies different types of care including choosing between cholinesterase inhibitors in AD versus antidepressant drugs in depression for instance or making a genetic diagnosis for FTD. Another challenging situation can be the disentanglement of PPA variants which all include predominant language impairment but are associated to variable neuropathological lesions [32]. SD could be distinguished from lv-PPA with an accuracy of 77%. As expected, the classifier, as well as the neuroradiologists, performed better on dementia known to have a strongly specific atrophy pattern (such as SD or FTD) [7] and worse on dementia with less specific atrophy patterns (LBD, CBD) [33,34]. Interestingly, the classifier allowed to distinguish SCD from the vast majority of neurodegenerative diseases with high accuracy. One can note that it performed better for SCD than for depression. One explanation could be the atrophy usually described in depression [35].

Compared to our SVM classifier, univariate classification based on AVS performed poorly. When analyzing the accuracy for diagnosis based on each of the volumes obtained with AVS, they ranged between 53% and 84%. With hippocampus alone, classifying rates rarely exceeded 70%, which is relatively low. In previous studies, the role of the hippocampus has been mainly evaluated for the diagnosis of AD versus controls or in mild cognitive impairment (MCI) populations to identify patients who will later progress to AD [8,9,11,36,37]. In our study, we evaluated MRI measurements in AD versus other dementia (FTD for instance), where hippocampal volumetry alone is known to perform poorly [38,39].

Poor performance of univariate classification and improvement when using SVM classification of both AVS volumes (balanced accuracy ranging from 60 to 80%) emphasize the fact that atrophy in dementia involves complex distributed spatial pattern. The only study comparing univariate (hippocampus) and multivariate analysis in two AVS (NeuroQuant™ and Neuroreader™) found different conclusions [13]. They didn't find any additional prognostic performance with multivariate analysis compared to univariate.

Nevertheless, this study focused on prediction of progression to AD among MCI patients, an objective that differs from ours. Finally, the SVM classifier using whole gray matter generally performed better than the multivariate analyses of both AVS. This is likely because the pattern of atrophy may not coincide with the boundaries of the anatomical regions delineated by AVS. This demonstrates the interest of letting the algorithm learn a discriminative pattern from the whole gray matter, without prior, rather using anatomical boundaries provided by AVS.

Neuroradiological classification was generally more accurate than hippocampal volumetry using AVS. The only exception was for LBD vs LOAD, a differential diagnosis for which anatomical MRI does not bring much relevant information and for which all approaches performed relatively poorly. Neuroradiological classification and SVM-WGM generally achieved similar performance. Nevertheless, the performance of SVM-WGM was superior for EOAD vs FTD. This indicates that an automatic classifier can be a useful tool to assist trained neuroradiologists for difficult situations.

Our study also demonstrates the feasibility of those techniques in the context of routine MRI data of varying image quality and acquired at different magnetic field strength. AVS segmentation and SVM classification were successful on almost every MRI.

One limitation of our study is the use of a binary classifier which does not totally correspond to the clinical practice where patients can have multiple diagnostic hypotheses. Further investigations could include multi-group classification instead of paired groups, in order to obtain a probability related to each potential diagnosis. Another limitation that we did not include healthy controls but rather used two control groups composed of patients with depression and SCD respectively. However, this situation is representative of the clinical routine: patients seen in a memory clinic are usually diagnosed with a neurological or a psychiatric condition, or present with subjective cognitive impairment, and are thus not "pure" control subjects.

As AVS start to be implemented in clinical routine, a final step in the analysis of raw AVS volumes could be a classification with an SVM based on all the AVS data. By analogy with AVS, our SVM-WGM classifier could be implemented in the post-processing of MRI in clinical routine. Thus, neuroradiologists could use the indication provided by the automatic classifier to refine their diagnosis. Also, in our study, neuroradiologists were operating in highly specialized centers and had considerable experience with different types of dementia (including rare diseases). It is thus conceivable that an automatic classifier would be of even greater help in less specialized centers.

**Conclusion**

Our study supports the applicability of computer-assisted diagnostic tools such as AVS and SVM classifiers to clinical routine data. When facing various dementia disorders, the accuracy of univariate volumetric analysis is too low to assist clinical decision making. In a clinical routine setting, automatic classifiers provide high diagnostic accuracy for distinguishing between several types of dementia. The implementation of advanced MRI-based computer-assisted diagnostic tools in clinical routine, such as SVM classification, could help to improve diagnostic accuracy.

**Table 1. Demographic and clinical characteristics of the population.**

Group differences were assessed with ANOVA for continuous variables and $X^2$ test for discrete variables. Data are expressed as mean +/- SD.

CBD = Cortico-basal syndrome, Depr. = Depression, EarlyAD = Early-onset AD, FTD = Fronto-temporal dementia of the behavioral type, LBD= Lewy body dementia, LateAD = Late-Onset-AD, lv-PPA = logopenic variant of Primary progressive aphasia, SCD= Subjective Cognitive Decline, SD = Semantic variant of primary progressive aphasia

| Diagnosis | Number | Age mean ± SD [range] | Gender | MMSE mean ± SD [range] | Magnetic Field (1T / 1.5T / 3T) |
|---|---|---|---|---|---|
| CBD | 31 | 69.75 ± 1.4 | 16 M / 15 F | 23.2 ± 4.5 | 16 / 10 / 5 |
| Depression | 24 | 64.52 ± 1.6 | 6 M / 18 F | 25.2 ± 3.2 | 18 / 3 / 3 |
| EarlyAD | 34 | 59.72 ± 1.49 | 13 M / 21 F | 20.0 ± 5.5 | 21 / 7 / 6 |
| FTD | 39 | 67.31 ± 1.315 | 22 M / 17 F | 23.2 ± 4.2 | 19 / 7 / 13 |
| LBD | 22 | 70.6 ± 1.75 | 13 M / 9 F | 22.3 ± 6.1 | 13 / 5 / 4 |
| LateAD | 49 | 73.5 ± 1.21 | 25 M / 24 F | 22.4 ± 4.1 | 24 / 9 / 16 |
| lv-PPA | 23 | 67.05 ± 1.7 | 15 M / 8 F | 19.9 ± 5.2 | 6 / 6 / 11 |
| SD | 17 | 65.36 ± 1.99 | 10 M / 7 F | 20.9 ± 8.1 | 7 / 5 / 5 |
| SCD | 12 | 72.5 ± 2.15 | 3 M / 9 F | 25,2 ± 2.9 | 3/3/6 |
| p-value | | < 0.0001 | 0.25 | 0.01 | 0.12 |

**Table 2. Classification results for univariate classification from hippocampal volumes obtained with Neuroreader™ ASS.** For each pair of possible diagnoses, we report the balanced accuracy. Chance level classification is at 50%. Colder colors (green/blue) correspond to less accurate classifications while warmer colors (red/orange) correspond to more accurate classifications.

CBD = Cortico-basal syndrome, Depr. = Depression, EarlyAD = Early-onset AD, FTD = Fronto-temporal dementia of the behavioral type, LBD= Lewy body dementia, LateAD = Late-Onset-AD, lv-PPA = logopenic variant of Primary progressive aphasia, SCD= Subjective Cognitive Decline, SD = Semantic variant of primary progressive aphasia

|        | SCD | Depr. | EarlyAD | LateAD | CBD | LBD | FTD | lv-PPA | SD  |
|--------|-----|-------|---------|--------|-----|-----|-----|--------|-----|
| SCD    | X   | X     | 53%     | 65%    | 56% | 55% | 63% | 49%    | 64% |
| Depr.  | X   | X     | 61%     | 71%    | 53% | 59% | 70% | 60%    | 71% |
| EarlyAD| 53% | 61%   | X       | 58%    | 59% | 48% | 60% | 52%    | 63% |
| LateAD | 65% | 71%   | 58%     | X      | 66% | 62% | 48% | 69%    | 46% |
| CBD    | 56% | 53%   | 59%     | 66%    | X   | 60% | 66% | 57%    | 68% |
| LBD    | 55% | 59%   | 48%     | 62%    | 60% | X   | 58% | 53%    | 60% |
| FTD    | 63% | 70%   | 60%     | 48%    | 66% | 58% | X   | 66%    | 52% |
| lv-PPA | 49% | 60%   | 52%     | 69%    | 57% | 53% | 66% | X      | 70% |
| SD     | 64% | 71%   | 63%     | 46%    | 68% | 60% | 52% | 70%    | X   |

**Table 3. Classification results for SVM classification from Whole Gray Matter maps.** For each pair of possible diagnoses, we report the balanced accuracy. Chance level is at 50%. Colder colors (green/blue) correspond to less accurate classifications while warmer colors (red/orange) correspond to more accurate classifications.

CBD = Cortico-basal syndrome, Depr. = Depression, EarlyAD = Early-onset AD, FTD = Fronto-temporal dementia of the behavioral type, LBD= Lewy body dementia, LateAD = Late-Onset-AD, lv-PPA = logopenic variant of Primary progressive aphasia, SCD= Subjective Cognitive Decline, SD = Semantic variant of primary progressive aphasia

|        | SCD | Depr. | EarlyAD | LateAD | CBD | LBD | FTD | lv-PPA | SD  |
|--------|-----|-------|---------|--------|-----|-----|-----|--------|-----|
| SCD    | X   | X     | 90%     | 85%    | 87% | 69% | 80% | 75%    | 87% |
| Depr.  | X   | X     | 83%     | 73%    | 78% | 71% | 82% | 66%    | 86% |
| EarlyAD| 90% | 83%   | X       | 59%    | 70% | 77% | 82% | 67%    | 71% |
| LateAD | 85% | 73%   | 59%     | X      | 78% | 52% | 74% | 54%    | 73% |
| CBD    | 87% | 78%   | 70%     | 78%    | X   | 55% | 67% | 58%    | 88% |
| LBD    | 69% | 71%   | 77%     | 52%    | 55% | X   | 67% | 54%    | 84% |
| FTD    | 80% | 82%   | 82%     | 74%    | 67% | 67% | X   | 70%    | 73% |
| lPPA   | 75% | 66%   | 67%     | 54%    | 58% | 54% | 70% | X      | 77% |
| SD     | 87% | 86%   | 71%     | 73%    | 88% | 84% | 73% | 77%    | X   |

**Table 4. Comparative performances of Neuroradiologists, univariate AVS, and automatic classifiers.**

The three diagnostic classification tasks are Depression vs LOAD,

FTD vs EOAD and LBD vs LOAD.

AVS = Automated Volumetry Software

SVM-AVS = Support Vector Machine Automated Volumetry Software

|  | Depression vs LOAD | FTD vs EOAD | LBD vs LOAD |
|---|---|---|---|
| **Neuroradiologist 1** | 77% | 72% | 57% |
| **Neuroradiologist 2** | 72% | 75% | 63% |
| **Hippocampal volumetry (AVS)** | 71% | 60% | 62% |
| **SVM-AVS (VolBrain)** | 60% | 67% | 54% |
| **SVM-AVS (Neuroreader)** | 76% | 67% | 63% |
| **SVM-WGM** | 73% | 82% | 52% |

# Appendix

**Supplementary Figure 1.** Spatial pattern learned by the classification algorithm. The maps represent contribution of each voxel to classification towards a given class (blue/green) or the other (yellow/red). Left panel: FTD (in yellow/red) vs. EarlyAD (in blue/green) displaying an anteroposterior gradient of atrophy. Right: LateAD (in blue/green) vs. depression (in yellow red) with medial temporal lobe voxels mostly blue/green.

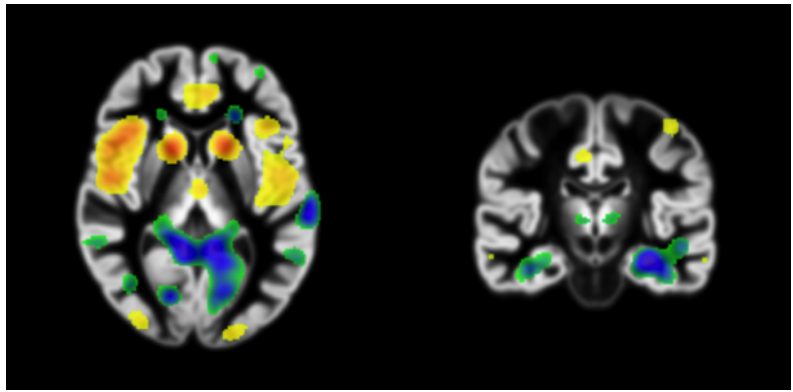

**Supplementary Table 1.** Patients Flow Chart

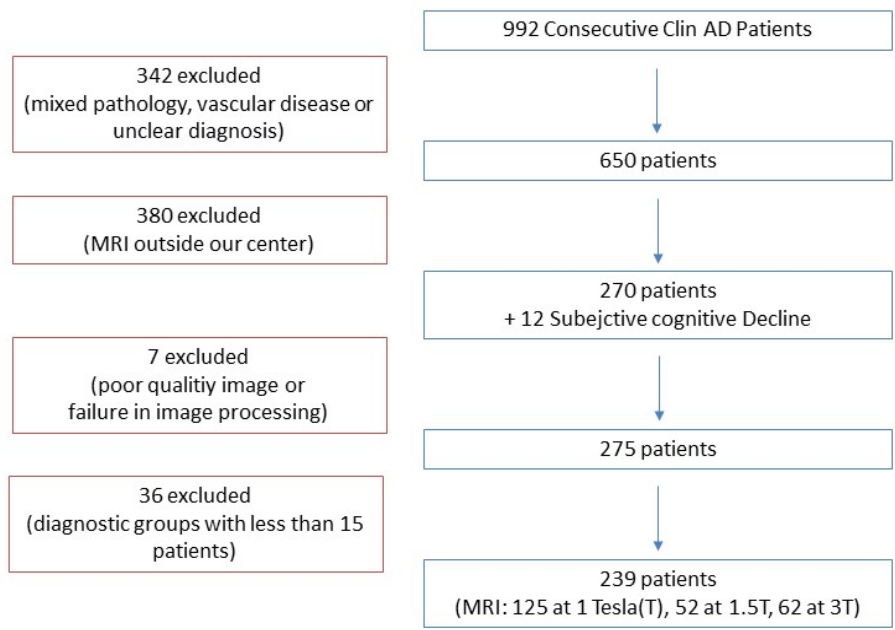

**Supplementary Table 2.** Patients MRI Sequence parameters.

Magnetic Field Strength = MF, T = Tesla, TE = Echo time, TR = Repetition time, TI = Inversion Time, ST = Slice Thickness, FA = Flip Angle, NA = Number of averages, PS = Pixel spacing, MT = Matrix Type

| MRI machine/ Sequence type | | MF (T) | TE Min (ms) | TE Max (ms) | TR Min (ms) | TR Max (ms) | TI Min (ms) | TI Max (ms) | ST Min (mm) | ST Max (mm) | FA Min (°) | FA Max (°) | NA Min | NA Max | PS Min (mm) | PS Max (mm) | MT 256X 192 | MT 192X 192 | MT 256X 256 | MT 224X 224 | MT 288X 288 | MT 288X 224 |
|---|---|---|---|---|---|---|---|---|---|---|---|---|---|---|---|---|---|---|---|---|---|---|
| Philips Panorama HFO | 123 | | | | | | | | | | | | | | | | | | | | | |
| eT1W 3D TFE | 123 | 1 | 3.90 | 4.35 | 8.33 | 9.08 | NA | NA | 1.20 | 1.20 | 8 | 8 | 1 | 1 | 0.99 | 1 | NA | NA | NA | NA | NA | NA |
| GE SIGNA HD | 63 | | | | | | | | | | | | | | | | | | | | | |
| N/A | 1 | 3 | NA | NA | NA | NA | NA | NA | NA | NA | NA | NA | NA | NA | NA | NA | NA | NA | NA | NA | NA | NA |
| 3D T1 BRAVO | 62 | 3 | 2.52 | 3.57 | 6.24 | 9.08 | 380 | 450 | 1 | 1.40 | 11 | 15 | 0.50 | 1 | 0.49 | 1 | 192 | 192 | 38 | 1 | 14 | 9 |
| GE SIGNA OPTIMA | 9 | | | | | | | | | | | | | | | | | | | | | |
| FSPGR 3D | 9 | 1.5 | 1.97 | 2.58 | 6.19 | 8.04 | 300 | 450 | 1 | 1.20 | 12 | 15 | 0.90 | 1 | 0.47 | 0.94 | | | | 9 | | |
| GE SIGNA EXCITE | 44 | | | | | | | | | | | | | | | | | | | | | |
| 3D T1 | 28 | 1.5 | 1.88 | 8 | 8.89 | 23 | 600 | 600 | 1.40 | 1.50 | 10 | 35 | 0.75 | 1 | 0.86 | 0.94 | | 28 | | | | |
| 3D IR-FSPGR | 16 | 1.5 | 1.87 | 2.56 | 6.58 | 9.80 | 600 | 1000 | 1.40 | 1.50 | 8 | 10 | 1 | 1 | 0.94 | 1.02 | 13 | | 3 | | | |

**Supplementary Table 3.** Mean volumes obtained through automatical segmentation using Neuroreader™. Volumes are expressed in $cm^3$ or as a percentage of Total Intracranial Volume. P-value were calculated using an ANOVA.

CBD = Cortico-basal degeneration, EOAD = Early-onset AD, FTD = Fronto-temporal dementia of the behavioral type, LBD= Lewy body dementia, LOAD = Late-Onset-AD, lv-PPA = logopenic variant of Primary progressive aphasia, SD = Semantic dementia, GM = Grey Matter, WM = White Matter, CSF = Cerebrospinal Fluid

| | | CBD | Depr. | EarlyAD | FTD | LBD | LateAD | lv-PPA | SD | SCD | p |
|---|---|---|---|---|---|---|---|---|---|---|---|
| Pallidum (Vol/TIV) | | 0,01 | 0,01 | 0,01 | 0,00 | 0,01 | 0,00 | 0,01 | 0,01 | 0,01 | < 0,0001 |
| | SD | 0,13 | 0,14 | 0,16 | 0,12 | 0,15 | 0,14 | 0,14 | 0,14 | 0,17 | |
| Temp. L.(vol/TIV) | | 200,20 | 195,91 | 186,18 | 187,04 | 197,12 | 189,07 | 190,69 | 168,77 | 188,06 | 0,01 |
| | SD | 4,55 | 5,17 | 4,41 | 4,05 | 5,40 | 3,62 | 5,28 | 6,14 | 7,31 | |
| Occip. L. (Vol/TIV) | | 91,35 | 89,01 | 87,09 | 89,47 | 94,32 | 89,07 | 90,11 | 92,43 | 87,18 | 0,49 |
| | SD | 2,10 | 2,39 | 2,03 | 1,87 | 2,49 | 1,67 | 2,44 | 2,83 | 3,37 | |
| Pariet. L. (ml) | | 176,31 | 181,82 | 171,61 | 175,42 | 187,84 | 178,10 | 172,76 | 188,09 | 175,50 | 0,07 |
| | SD | 3,81 | 4,33 | 3,69 | 3,39 | 4,52 | 3,03 | 4,42 | 5,14 | 6,12 | |
| Front. L. (ml) | | 348,97 | 357,97 | 349,06 | 310,57 | 369,23 | 351,99 | 352,31 | 359,63 | 338,87 | < 0,0001 |
| | SD | 7,97 | 9,06 | 7,73 | 7,11 | 9,46 | 6,34 | 9,26 | 10,77 | 12,82 | |
| Thalamus (Vol/TIV) | | 0,02 | 0,02 | 0,02 | 0,02 | 0,02 | 0,01 | 0,02 | 0,03 | 0,03 | 0,03 |
| | SD | 0,87 | 0,94 | 0,91 | 0,85 | 0,90 | 0,88 | 0,92 | 0,91 | 0,91 | |
| Putamen (Vol/TIV) | | 0,01 | 0,01 | 0,01 | 0,01 | 0,02 | 0,01 | 0,02 | 0,02 | 0,02 | < 0,0001 |
| | SD | 0,35 | 0,43 | 0,40 | 0,35 | 0,41 | 0,41 | 0,41 | 0,39 | 0,46 | |
| Ventric. (Vol/TIV) | | 0,21 | 0,24 | 0,21 | 0,19 | 0,25 | 0,17 | 0,25 | 0,29 | 0,34 | 0,01 |
| | SD | 2,79 | 1,58 | 2,14 | 2,75 | 2,71 | 2,31 | 2,39 | 2,55 | 2,15 | |
| Caud. N. (Vol/TIV) | | 0,01 | 0,01 | 0,01 | 0,01 | 0,01 | 0,01 | 0,01 | 0,02 | 0,02 | 0,00 |
| | SD | 0,34 | 0,37 | 0,34 | 0,30 | 0,37 | 0,36 | 0,35 | 0,36 | 0,40 | |
| Amygd. (vol/TIV) | | 0,00 | 0,01 | 0,00 | 0,00 | 0,00 | 0,00 | 0,01 | 0,01 | 0,01 | < 0,0001 |
| | SD | 0,09 | 0,11 | 0,09 | 0,09 | 0,09 | 0,09 | 0,10 | 0,08 | 0,12 | |
| Hippoc. (Vol/TIV) | | 0,01 | 0,01 | 0,01 | 0,01 | 0,01 | 0,01 | 0,01 | 0,01 | 0,02 | 0,00 |
| | SD | 0,36 | 0,38 | 0,34 | 0,33 | 0,35 | 0,33 | 0,36 | 0,33 | 0,36 | |
| CSF (ml) | | 15,32 | 17,42 | 14,85 | 13,66 | 18,19 | 12,19 | 17,79 | 20,69 | 24,63 | 0,03 |
| | SD | 538,57 | 473,77 | 496,25 | 527,49 | 494,12 | 490,73 | 502,12 | 495,23 | 431,50 | |
| GM (ml) | | 0,62 | 0,70 | 0,68 | 0,61 | 0,65 | 0,64 | 0,67 | 0,63 | 0,68 | 0,03 |
| | SD | 433,49 | 495,61 | 414,47 | 398,66 | 487,03 | 441,10 | 438,98 | 438,06 | 468,69 | |
| WM (ml) | | 20,17 | 22,92 | 19,55 | 17,98 | 23,94 | 16,04 | 23,41 | 27,23 | 32,41 | 0,41 |
| | SD | 587,40 | 524,22 | 572,60 | 559,97 | 567,26 | 565,41 | 573,40 | 580,83 | 506,32 | |

**Supplementary Table 4.** Classification results for univariate classification from gray matter volumes obtained using Neuroreader™. For each pair of possible diagnoses, we report the balanced accuracy. Chance level is at 50%. Colder colors (green/blue) correspond to less accurate classifications while warmer colors (red/orange) correspond to more accurate classifications.

|        | SCD | Depr. | EarlyAD | LateAD | CBD | LBD | FTD | lv-PPA | SD  |
|--------|-----|-------|---------|--------|-----|-----|-----|--------|-----|
| SCD    | X   | X     | 62%     | 61%    | 61% | 59% | 66% | 58%    | 54% |
| Depr.  | X   | X     | 66%     | 66%    | 65% | 63% | 74% | 62%    | 62% |
| EarlyAD| 62% | 66%   | X       | 50%    | 52% | 56% | 60% | 55%    | 46% |
| LateAD | 61% | 66%   | 50%     | X      | 47% | 55% | 64% | 47%    | 50% |
| CBD    | 61% | 65%   | 52%     | 47%    | X   | 55% | 63% | 56%    | 46% |
| LBD    | 59% | 63%   | 56%     | 55%    | 55% | X   | 67% | 55%    | 52% |
| FTD    | 66% | 74%   | 60%     | 64%    | 63% | 67% | X   | 70%    | 64% |
| lv-PPA | 58% | 62%   | 55%     | 47%    | 56% | 55% | 70% | X      | 55% |
| SD     | 54% | 62%   | 46%     | 50%    | 46% | 52% | 64% | 55%    | X   |

**Supplementary Table 5.** Classification results for univariate classification from caudate nucleus volumes obtained using Neuroreader™. For each pair of possible diagnoses, we report the balanced accuracy. Chance level is at 50%. Colder colors (green/blue) correspond to less accurate classifications while warmer colors (red/orange) correspond to more accurate classifications.

|        | SCD | Depr. | EarlyAD | LateAD | CBD | LBD | FTD | lv-PPA | SD  |
|--------|-----|-------|---------|--------|-----|-----|-----|--------|-----|
| SCD    | X   | X     | 78%     | 68%    | 79% | 65% | 87% | 70%    | 69% |
| Depr.  | X   | X     | 67%     | 54%    | 66% | 55% | 72% | 58%    | 57% |
| EarlyAD| 78% | 67%   | X       | 61%    | 51% | 56% | 57% | 67%    | 55% |
| LateAD | 68% | 54%   | 61%     | X      | 60% | 50% | 67% | 51%    | 53% |
| CBD    | 79% | 66%   | 51%     | 60%    | X   | 58% | 62% | 64%    | 55% |
| LBD    | 65% | 55%   | 56%     | 50%    | 58% | X   | 66% | 51%    | 51% |
| FTD    | 87% | 72%   | 57%     | 67%    | 62% | 66% | X   | 69%    | 61% |
| lv-PPA | 70% | 58%   | 67%     | 51%    | 64% | 51% | 69% | X      | 48% |
| SD     | 69% | 57%   | 55%     | 53%    | 55% | 51% | 61% | 48%    | X   |

**Supplementary Table 6.** Classification results for univariate classification from amygdala volumes obtained using Neuroreader™. For each pair of possible diagnoses, we report the balanced accuracy. Chance level is at 50%. Colder colors (green/blue) correspond to less accurate classifications while warmer colors (red/orange) correspond to more accurate classifications.

|        | SCD | Depr. | EarlyAD | LateAD | CBD | LBD | FTD | lv-PPA | SD  |
|--------|-----|-------|---------|--------|-----|-----|-----|--------|-----|
| SCD    | X   | X     | 60%     | 69%    | 70% | 65% | 67% | 62%    | 66% |
| Depr.  | X   | X     | 62%     | 67%    | 67% | 62% | 72% | 55%    | 76% |
| EarlyAD| 60% | 62%   | X       | 52%    | 56% | 52% | 53% | 59%    | 59% |
| LateAD | 69% | 67%   | 52%     | X      | 58% | 56% | 50% | 62%    | 58% |
| CBD    | 70% | 67%   | 56%     | 58%    | X   | 46% | 59% | 63%    | 65% |
| LBD    | 65% | 62%   | 52%     | 56%    | 46% | X   | 57% | 62%    | 64% |
| FTD    | 67% | 72%   | 53%     | 50%    | 59% | 57% | X   | 65%    | 58% |
| lv-PPA | 62% | 55%   | 59%     | 62%    | 63% | 62% | 65% | X      | 71% |
| SD     | 66% | 76%   | 59%     | 58%    | 65% | 64% | 58% | 71%    | X   |

**Supplementary Table 7.** Classification results for univariate classification from temporal lobe volumes obtained using Neuroreader™. For each pair of possible diagnoses, we report the balanced accuracy. Chance level is at 50%. Colder colors (green/blue) correspond to less accurate classifications while warmer colors (red/orange) correspond to more accurate classifications.

|        | SCD | Depr. | EarlyAD | LateAD | CBD | LBD | FTD | lv-PPA | SD  |
|--------|-----|-------|---------|--------|-----|-----|-----|--------|-----|
| SCD    | X   | X     | 59%     | 64%    | 52% | 59% | 67% | 59%    | 74% |
| Depr.  | X   | X     | 63%     | 64%    | 62% | 63% | 69% | 59%    | 82% |
| EarlyAD| 59% | 63%   | X       | 51%    | 54% | 48% | 55% | 53%    | 64% |
| LateAD | 64% | 64%   | 51%     | X      | 60% | 49% | 57% | 52%    | 61% |
| CBD    | 52% | 62%   | 54%     | 60%    | X   | 55% | 67% | 57%    | 73% |
| LBD    | 59% | 63%   | 48%     | 49%    | 55% | X   | 62% | 49%    | 75% |
| FTD    | 67% | 69%   | 55%     | 57%    | 67% | 62% | X   | 52%    | 61% |
| lv-PPA | 59% | 59%   | 53%     | 52%    | 57% | 49% | 52% | X      | 61% |
| SD     | 74% | 82%   | 64%     | 61%    | 73% | 75% | 61% | 61%    | X   |

**Supplementary Table 8.** Classification results for univariate classification from frontal lobe volumes obtained using Neuroreader™. For each pair of possible diagnoses, we report the balanced accuracy. Chance level is at 50%. Colder colors (green/blue) correspond to less accurate classifications while warmer colors (red/orange) correspond to more accurate classifications.

|        | SCD | Depr. | EarlyAD | LateAD | CBD | LBD | FTD | lv-PPA | SD  |
|--------|-----|-------|---------|--------|-----|-----|-----|--------|-----|
| SCD    | X   | X     | 47%     | 57%    | 62% | 50% | 76% | 55%    | 59% |
| Depr.  | X   | X     | 58%     | 61%    | 69% | 63% | 82% | 62%    | 61% |
| EarlyAD| 47% | 58%   | X       | 56%    | 62% | 48% | 71% | 59%    | 58% |
| LateAD | 57% | 61%   | 56%     | X      | 59% | 53% | 74% | 52%    | 55% |
| CBD    | 62% | 69%   | 62%     | 59%    | X   | 68% | 72% | 55%    | 54% |
| LBD    | 50% | 63%   | 48%     | 53%    | 68% | X   | 80% | 58%    | 54% |
| FTD    | 76% | 82%   | 71%     | 74%    | 72% | 80% | X   | 75%    | 73% |
| lv-PPA | 55% | 62%   | 59%     | 52%    | 55% | 58% | 75% | X      | 45% |
| SD     | 59% | 61%   | 58%     | 55%    | 54% | 54% | 73% | 45%    | X   |

**Supplementary Table 9.** Classification results for univariate classification from parietal lobe volumes obtained using Neuroreader™. For each pair of possible diagnoses, we report the balanced accuracy. Chance level is at 50%. Colder colors (green/blue) correspond to less accurate classifications while warmer colors (red/orange) correspond to more accurate classifications.

|        | SCD | Depr. | EarlyAD | LateAD | CBD | LBD | FTD | lv-PPA | SD  |
|--------|-----|-------|---------|--------|-----|-----|-----|--------|-----|
| SCD    | X   | X     | 73%     | 62%    | 72% | 58% | 74% | 74%    | 59% |
| Depr.  | X   | X     | 71%     | 70%    | 75% | 59% | 74% | 72%    | 61% |
| EarlyAD| 73% | 71%   | X       | 58%    | 52% | 61% | 54% | 50%    | 58% |
| LateAD | 62% | 70%   | 58%     | X      | 60% | 57% | 62% | 61%    | 56% |
| CBD    | 72% | 75%   | 52%     | 60%    | X   | 62% | 54% | 52%    | 57% |
| LBD    | 58% | 59%   | 61%     | 57%    | 62% | X   | 66% | 67%    | 46% |
| FTD    | 74% | 74%   | 54%     | 62%    | 54% | 66% | X   | 48%    | 59% |
| lv-PPA | 74% | 72%   | 50%     | 61%    | 52% | 67% | 48% | X      | 59% |
| SD     | 59% | 61%   | 58%     | 56%    | 57% | 46% | 59% | 59%    | X   |

**Supplementary Table 10.** Classification results for SVM classification from fall volumes obtained using volBrain (on top) and Neuroreader™ (at the bottom). For each pair of possible diagnoses, we report the balanced accuracy. Chance level is at 50%. Colder colors (green/blue) correspond to less accurate classifications while warmer colors (red/orange) correspond to more accurate classifications.

**VolBrain**

|        | SCD | Depr. | EarlyAD | LateAD | CBD | LBD | FTD | lv-PPA | SD  |
|--------|-----|-------|---------|--------|-----|-----|-----|--------|-----|
| SCD    | X   | X     | 82%     | 57%    | 81% | 64% | 80% | 60%    | 72% |
| Depr.  | X   | X     | 71%     | 68%    | 68% | 70% | 79% | 72%    | 85% |
| EarlyAD| 82% | 71%   | X       | 72%    | 65% | 68% | 73% | 52%    | 58% |
| LateAD | 57% | 68%   | 72%     | X      | 78% | 68% | 77% | 68%    | 62% |
| CBD    | 81% | 68%   | 65%     | 78%    | X   | 60% | 56% | 59%    | 67% |
| LBD    | 64% | 70%   | 68%     | 68%    | 60% | X   | 69% | 56%    | 77% |
| FTD    | 80% | 79%   | 73%     | 77%    | 56% | 69% | X   | 60%    | 54% |
| lv-PPA | 60% | 72%   | 52%     | 68%    | 59% | 56% | 60% | X      | 71% |
| SD     | 72% | 85%   | 58%     | 62%    | 67% | 77% | 54% | 71%    | X   |

**NeuroReader**

|        | SCD | Depr. | EarlyAD | LateAD | CBD | LBD | FTD | lv-PPA | SD  |
|--------|-----|-------|---------|--------|-----|-----|-----|--------|-----|
| SCD    | X   | X     | 62%     | 54%    | 78% | 63% | 74% | 62%    | 79% |
| Depr.  | X   | X     | 65%     | 60%    | 75% | 56% | 77% | 70%    | 79% |
| EarlyAD| 62% | 65%   | X       | 61%    | 80% | 70% | 67% | 62%    | 73% |
| LateAD | 54% | 60%   | 61%     | X      | 69% | 54% | 66% | 64%    | 70% |
| CBD    | 78% | 75%   | 80%     | 69%    | X   | 63% | 60% | 60%    | 83% |
| LBD    | 63% | 56%   | 70%     | 54%    | 63% | X   | 69% | 54%    | 84% |
| FTD    | 74% | 77%   | 67%     | 66%    | 60% | 69% | X   | 65%    | 82% |
| lv-PPA | 62% | 70%   | 62%     | 64%    | 60% | 54% | 65% | X      | 65% |
| SD     | 79% | 79%   | 73%     | 70%    | 83% | 84% | 82% | 65%    | X   |